**Time-independent pricing of options in range bound markets**

Ovidiu Racorean

e-mail: decontatorul@hotmail.com

Abstract

Assuming that price of the underlying stock is moving in range bound, the Black-Scholes formula for options pricing supports a separation of variables. The resulting time-independent equation is solved employing different behavior of the option price function and three significant results are deduced. The first is the probability of stock price penetration through support or resistance level, called transmission coefficient. The second is the distance that price will go through once stock price penetrates out of the range bound. The last one is a predicted short time dramatic fall in the stock volatility right ahead of price tunneling. All three results are useful tools that give market practitioners valuable insights in choosing the right time to get involved in an option contract, about how far the price will go in case of a breakout, and how to correctly interpret volatility downfalls.

1. **Introduction**

The value of an option contract for an underlying stock having the price moving in range bound is not changing much, in time. More, along with time the option value erodes. Traders and investors that are cough in an option contract for such underlying stock are likely to lose their initial investment.

The moment of getting involved into an option contract must be carefully examined since is crucial for the future evolution of the option value.

In particular, for American options, choosing the right moment for buying an option contract gain more weight in the process of making trading decisions since it can be exercise at any time up to the date of expiration. Market practitioners could gain fast profits only by picking the right time for achieving option contacts which, in context of a range bound market, means to find the best odds in favor of a break-out through support/resistance levels to occur.

First step in attempting to find the probability of the penetration out of support/resistance levels is to consider a separation of variables in Black-Scholes valuation formula. The separation of variables is possible due to the stationary price moving in the region bounded by support and resistance levels.

This angle of viewing the Black-Scholes formula leads to a system of two equations, both of them equated with the same **λ** constant; one expressing the time evolution of option value, and one time-independent formula which constitutes the key element in deriving the further mathematical results.

Investigating the time-independent equation mentioned above one can emphasize that it represent the one dimensional Schrodinger equation for a particle in a box. This result should not be a surprise since the move of stock price in range bound, between support/resistance levels, is an exact replica of a particle moving in a box. Just like the particle could tunnel out of the box, the stock price can also break-out from de region bounded by the support and resistance levels. The tunneling of stock price may sound a little peculiar but is a phenomenon very familiar to traders that encounters this kind of price move almost every day.

The probability of the particle tunneling out of the box – also called transmission coefficient- is becoming, in the case of stock market, the probability of stock price to penetrate out of the support and resistance region.

Solving the time-independent equation, for stock price, in the particle in a box (physical) manner, leads naturally, to the transmission coefficient for the stock price to penetrate beyond the level of support or resistance.

Choosing an option contract having the strike price as nearest as possible to resistance or support, it can be noticed that transmission coefficient is depending only on interest rate, volatility and strike price of the underlying stock.

Using the transmission coefficient, market practitioners could get valuable insights about the right moment to buy an option contract of one underlying stock or another. The most probable tunneling stock price should be chosen as a basis for the option contract.

Although, the transmission coefficient is derived for the options market framework, it can also be used in other markets that exhibits a range bound behavior.

An extended discussion is dedicated to **λ** constant, whose values dictate the transmission coefficient applicability. It seems that only stocks having prices ranging in an interval from 2 USD to 15 USD can be subject of time-independent formula. Also, from values of **λ** another important result is derived; once the stock price penetrates the wall, the magnitude of the move can be anticipated knowing the **λ** constant value. The distance of break out through support/resistance wall is very important in knowing what to expect from stock price tunneling.

A third important result deduced from the time-independent formula is that of a predicted dramatic short term fall in the stock volatility right ahead of a price tunneling.

## 2. Separation of variables for Black-Scholes equation

The starting point of the analysis is the Black-Scholes equation which can be written in the form:

$$\frac{\partial w}{\partial t} = -\frac{1}{2}\sigma^2 S^2 \frac{\partial^2 w}{\partial S^s} - rS \frac{\partial w}{\partial S} + rw, \tag{1}$$

in the usual notations.

Using the method of separation of variables, solving equation (1) reduces to find the solution $w_{(S,t)}$ that is a product of two functions with only one variable:

$$w_{(S,t)} = \phi_{(S)} \varphi_{(t)} \tag{2}$$

in this case time and stock price.

Putting this into Black-Scholes equation, dropping the partial derivative in favor of ordinary derivative and rearranging the terms yields:

$$\frac{d\varphi_{(t)}}{dt}\frac{1}{\varphi_{(t)}} = -\frac{1}{2}\sigma^2 S^2 \frac{d^2\phi_{(S)}}{dS^2}\frac{1}{\phi_{(S)}} - rS\frac{d\phi_{(S)}}{dS}\frac{1}{\phi_{(S)}} + r \tag{3}$$

It follows from this, that both sides of equation (3) must equal a constant:

$$\frac{d\varphi_{(t)}}{dt}\frac{1}{\varphi_{(t)}} = \lambda \tag{4}$$

$$-\frac{1}{2}\sigma^2 S^2 \frac{d^2\phi_{(S)}}{dS^2}\frac{1}{\phi_{(S)}} - rS\frac{d\phi_{(S)}}{dS}\frac{1}{\phi_{(S)}} + r = \lambda \tag{5}$$

with **λ** constant. The **λ** constant is central to this presentation, and the next section is devoted to deduce its value from the time dependency of the option price.

## 3. Time dependency of options price

It should be said in the first place that the **λ** value is circumstantial deduced only by logical assessments of market and macroeconomic factors affecting option price in time.

Equation (4) can be written:

$$\frac{d\varphi_{(t)}}{dt} = \lambda \varphi_{(t)} \tag{6}$$

and is trivial to be solved to give:

$$\varphi_{(t)} = e^{\lambda t} \tag{7}$$

This equation is revealing the time dependency of the option price. The practice of options trading shows that value of an option decrease as time passes and the expiration approaches, in which case the constant **λ** bust be negative.

It is also very well known that the price of options decays slower for the stocks with a high volatility and faster for the stocks with low volatility. The volatility term also incorporates the distance between strike price and market price. To illustrate this assessment, a simple example from trading options experience is representative:

-having a strike price of 2,50 USD , the market price at 2,40 USD give a $\sigma$ of , say 130%. To take the example to the extreme, a strike price of 7,40 USD for the same market price, have $\sigma$ of 285%. It is easy to see that option value in the last case will decay more rapidly than in the first case.

Other factor that directly affects the price of options is the interest rate. The influence of the interest rate on the **λ** constant is not easy to quantify. It is taken into account, here, the influence that interest rate have on the stock price. An extended discussion will be in section 6.

Accounting to the above mentioned factors lead to the conclusion that **λ** constant must have the form:

$$\lambda = \frac{r}{\sigma} \tag{8}$$

The equation (7) should be modified as:

$$\varphi_{(t)} = e^{-\frac{r}{\sigma}t}, \tag{9}$$

which simply measures the rate of the decay in option price with time.

### 4. Stock prices moving in range bound

This section is dedicated to the part of the Black-Scholes formula that is not explicitly depending on time and we will write equation (5) in the form:

$$-\frac{1}{2}\sigma^2 S^2 \frac{d^2\phi_{(S)}}{dS^2} - rS\frac{d\phi_{(S)}}{dS} + r\phi_{(S)} = \lambda\phi_{(S)} \tag{10}$$

First focus the attention to the homogeneous part:

$$-\frac{1}{2}\sigma^2 S^2 \frac{d^2\phi_{(S)}}{dS^2} - rS\frac{d\phi_{(S)}}{dS} + r\phi_{(S)} = 0 \tag{11}$$

Equation (11) can be further written as:

$$\frac{d^2\phi_{(S)}}{dS^2} + \frac{2r}{\sigma^2 S}\frac{d\phi_{(S)}}{dS} - \frac{2r}{\sigma^2 S^2}\phi_{(S)} = 0 \tag{12}$$

In order to extract useful insights, the above equation will be put, using the notation $\ln\phi_{(S)} = \ln\psi_{(S)} - \frac{1}{2}\int \frac{2r}{\sigma^2 S}dS$, in the standard form:

$$\frac{d^2\psi_{(S)}}{dS^2} - \frac{r}{\sigma^2 S^2}\left(1 + \frac{r}{\sigma^2}\right)\psi_{(S)} = 0 \tag{13}$$

Rearranging the terms equation (13) takes the form:

$$-\frac{\sigma^4}{r(\sigma^2+r)}\frac{d^2\psi_{(S)}}{dS^2} + \frac{1}{S^2}\psi_{(S)} = 0 \tag{14}$$

The above equation is the time-independent Schrodinger equation for a zero energy particle, moving in a $V_{(S)} = \frac{1}{S^2}$ potential.

The shape of the potential can be seen in Figure 1, bellow:

Figure 1.

It should be noticed here the convention that will be used along this paper. The stock price at support level is set to be on the zeroth of the graph; the strike price of the option contract that is set to be on the resistance, noted with K on the Figure 1, represent the width of the region between support and resistance.

Throughout this paper, for the sake of simplicity, the mathematics will refer to call option contracts. However, the logic of put option contracts is similar.

The $V_0$ value is the potential for the resistance level price. Choosing the option having the strike price as nearest as possible to resistance the $V_0$ potential is:

$$V_0 = \frac{1}{K^2}, \tag{15}$$

where $K$ is the strike price for call option.

As the stock price takes higher values the potential $V_{(S)}$ decline but never reaches zero value and the wall width grows.

Notice from Figure 1 that in the absence of **λ** constant the price of underlying stock cannot breakout of the resistance level and is indefinitely trapped in a region bounded by resistance and support.

## 5. Time-independent option valuation formula

The conclusion that last section draws is that **λ** constant must not vanish in order for stock price to penetrate through support/resistance levels, and not remain indefinitely trapped in range bound.

Equating back, from (10) and (14) we have:

$$-\frac{\sigma^4}{r(\sigma^2+r)}\frac{d^2\psi_{(S)}}{dS^2} + \frac{1}{S^2}\psi_{(S)} = \lambda\psi_{(S)}, \tag{16}$$

which is easy to recognize, taking $V_{(S)} = \frac{1}{S^2}$ and rewriting (16),

$$-\frac{\sigma^4}{r(\sigma^2+r)}\frac{d^2\psi_{(S)}}{dS^2} + V_{(S)}\psi_{(S)} = \lambda\psi_{(S)}, \tag{17}$$

as being the Schrodinger equation of a particle with energy **λ** moving in a potential $V_{(S)}$.

It was showed in section 3 that $\lambda = \frac{r}{\sigma}$ and

$$-\frac{\sigma^4}{r(\sigma^2+r)}\frac{d^2\psi_{(S)}}{dS^2} + V_{(S)}\psi_{(S)} = \frac{r}{\sigma}\psi_{(S)}. \tag{18}$$

Solving the above equation give the evolution of option value as a function of the underlying stock price, in the absence of time component. This is a limit case, adapted for short time frames, still option valuation cannot be viewed independent from its time component in long time frames.

For a stock being in range bound, price is moving between resistance and support levels, and solving equation (18) reduces to find option price stationary solutions of the form:

$$\psi_{(S)} = \sqrt{\frac{2}{K}} \sin\frac{n\pi S}{K} \tag{19}$$

where K is the distance between resistance and support, as was already stated in the last section. Although is not the subject of this paper, it is not without important to mention here that having the stock price moving in range bound the option value seems to follow a dumping oscillator move, slowing its motion until it reach a stationary state at the maturity.

In Figure 2 are represented different levels of **λ** and how it affects the stock price movement.

Figure 2.

It can be noticed from Figure 2 that if $\lambda > V_0$ the market is trending, and not being in range bound the further discussion did not apply. In the further discussion, in the next section, it is assumed that $\lambda < V_0$.

## 6. The **λ** constant

The **λ** constant is equated, on the one hand with time decay of option price and on the other hand, with changes in option value along with underlying stock price evolution. The value of **λ** constant affects not only the time decay of option contracts, but also the stock market itself.

It can be seen in Figure 2 that the higher **λ** is, the tinnier the potential wall width and the lowest **λ** is, the bigger width of the wall will be. Accounting the two parameters **λ** consists of, interest rate r and volatility, it can be said that:

- The higher the interest rate or the lowest the volatility are, the tinnier the potential wall that stock price should tunnel through. Because the wall is tinny, even small moves in stock price could penetrate the potential wall. In periods of high interest rate big moves on the stock market should not be expected.
- If the interest rate is low or the volatility is high, the width of the wall grows and to penetrate the wall stock price should make big moves. In periods of low interested rate big moves should be expected in the market.

One crucial aspect that should be discuss here is the condition $\lambda < V_0$. Considering equation (15) and solving for the strike price, the above condition become:

$$K < \sqrt{\frac{1}{\lambda}} \qquad (19)$$

Remember that **λ** constant depends on interest rate $r$ and volatility $\sigma$. Taking the values of this two parameters to the extreme, say $r \in (0.01, 0.15)$ and $\sigma \in (0.02, 25)$, **λ** constant is varying, more or less, in the interval $\lambda \in (0.004, 10)$. According to inequality (19), the strike price must be situated in the interval:

$$K \in (1, 20)$$

It seems that transmission coefficient should be applied only for stocks having prices moving in a maximum range bound of 20 USD. The smallest the width of the range bound, the better the phenomenon is observed.

Once the stock price penetrates the wall, the magnitude of the move can be anticipated knowing the **λ** constant value. Figure 2 shows that at the boundary of the wall **λ** must equal the potential $V_r$, which means that, at this point, the stock price will be:

$$S_r = \sqrt{\frac{1}{\lambda}} \tag{20}$$

The penetration distance is then $d = S_r - K$. In other words, the price of underlying stock will tunnel through the wall until it reaches the $S_r$ price, penetrating through a distance:

$$d = \sqrt{\frac{\sigma}{r}} - K \tag{21}$$

To illustrate this important result, take the strike price $K$ of a certain stock at 2,40 USD, interest rate of 3% and volatility $\sigma$ at 47%, so that the **λ** will be 0,064. From eq. 20 is trivial to derive the stock price $S_r$ at the end of the tunnel as being 3,95 USD. The price will penetrate the distance 1,85.

The distance of break out through support/resistance wall is very important in knowing what to expect from stock price tunneling.

**7. Transmission coefficient**

Figure 3 shows the trajectory of a stock price tunneling the wall bounded by resistance level, where, the strike price is also situated.

Figure 3.

The price movement is fragmented in regions according to the behavior of the $\psi_{(S)}$ function. The time-independent formula will be further solved and discussed separately for every of this regions.

To the left of the barrier, in the Region I, bounded by the support and resistance levels, function is :

$$\psi_1 = Ae^{ikS} + Be^{-ikS} \tag{22}$$

with $k = \sqrt{\frac{r}{\sigma^4}(\sigma^2 + r)\lambda}$

Inside the barrier wall, in the region II, function is exponential:

$$\psi_2 = Ce^{qS} + De^{-qS} \tag{23}$$

with $q = \sqrt{\frac{r}{\sigma^4}(\sigma^2 + r)(V_{(S)} - \lambda)}$

To the right of the barrier, in the region III

$$\psi_3 = Fe^{ikS} \tag{24}$$

The probability T for the price passing the barrier is given by:

$$T = \frac{|F|^2}{|A|^2} \tag{25}$$

To find T, one simply have to find A and F. Those amplitudes can be found taken into account that both the function $\psi$ and its derivative must be continuous at the barrier points. A system of equation is derived from these boundary conditions and has been solved to give:

$$|A|^2 = |F|^2 \left( \frac{(k^2+q^2)^2}{4k^2q^2} \sinh^2(qd) + 1 \right) \tag{26}$$

Where d is the width of the potential wall.

Expressing k and q in terms of V and $\lambda$, transmission coefficient may be written:

$$T = \left( \frac{V^2}{4\lambda(V-\lambda)} \sinh^2(qd) + 1 \right)^{-1} \tag{27}$$

For qd >>1 the the equation (26) is dominated by the sinh term and can be approximated:

$$T = e^{-2qd} \tag{28}$$

Considering all the possibilities for the barrier potential:

$$T = e^{-2\sum_i (q_i d_i)} \tag{29}$$

and taking the integral instead of the sum:

$$T = e^{-2\sqrt{\frac{r}{\sigma^4}(\sigma^2+r)} \int_K^{S_1} \sqrt{\frac{1}{S^2} - \lambda}\, dS} \tag{30}$$

Evaluating the integral in (29), from strike price K to the price outside the wall $S_1$ rearranging the terms the probability of price tunneling the potential wall is:

$$T = e^{-2\sqrt{\frac{r}{\sigma^4}(\sigma^2+r)} \left[ \frac{1}{2}\ln\left( \left|\frac{\sqrt{1-\lambda K^2}+1}{\sqrt{1-\lambda K^2}-1}\right| \right) - \sqrt{1-\lambda K^2} \right]} \tag{31}$$

Since $\lambda = \frac{r}{\sigma}$, the transmission coefficient is:

$$T = e^{-2\sqrt{\frac{r}{\sigma^4}(\sigma^2+r)} \left[ \frac{1}{2}\ln\left( \left|\frac{\sqrt{1-\frac{r}{\sigma}K^2}+1}{\sqrt{1-\frac{r}{\sigma}K^2}-1}\right| \right) - \sqrt{1-\frac{r}{\sigma}K^2} \right]} \tag{32}$$

and represent the probability of stock price to penetrate through the walls defined by support or resistance level.

The transmission coefficient formula may look complicated, but is easy to be compute since it depends only on known parameters: interest rate, volatility and strike price.

Maintaining the same values of the strike price as in the latter examples, $K = 2,40$ USD, and varying the values of the other two parameters, different values of transmission coefficient and penetrating distances are shown in the Table1 and Table 2.

Table 1.

Table 2.

Notice that keeping volatility constant (Table 1), as the interest rate increase, the probability of tunneling is high due to smaller wall width of penetration.

It also can be notice that keeping interest rate constant (Table 2), as the volatility raise, the transmission coefficient diminishes due to higher width of the wall to be penetrated.

An extremely important result should be also mentioned here, because the interest rate remains unchanged during the life time of an option contract, in order to see a stock price tunneling effect, the potential wall must be thin, as result the volatility must be low. **Before a tunneling effect can occur in the stock market, a dramatic fall in the stock volatility, in a very short period of time, should be seen.**

Figure 4 captures tunneling effect of some stocks along with the predicted fall in the volatility.

Figure 4.

Both pictures are explicitly showed that the moment of dramatically fall in the volatility coincides with an abrupt rise in the stock price.

8. **Empirical tests**

In the everyday trading practice on the market, tunneling of stock price out of the range bound that it moved for some time, is a very common phenomenon. I guess there is no trader that shouldn't see, at least one time, a stock price effectively "exploding", rising very fast in a short period of time. Empirical evidence of stock tunneling could be find every day in the stock market, and not only. That makes this section purely informative. The interested reader can easily find many other examples on the financial market and compute transmission coefficient.

The Table 3 below gathers together some data for four different stocks, which were track from the site www.optionistics.com:

Table 3.

The table is instructive by showing the way K, d and T are computed. Notice, also, the important fall in the volatility ahead of price tunneling.

### 9. Conclusions and suggestions for further work

In the hypothesis of the underlying stock price of an option contract being in range bound the Black-Scholes equation can be solved by separation of variables. Separation of variables scheme applied to option valuation leads to a system of two equations, both of them being equated with the same $\lambda$ constant.

First equation represents the time decay of option value and is used to deduce the value of the **λ** constant.

The **λ** constant is essential for understanding further the concepts of the second equation which is time-independent and represents the option value dependency of the underlying stock price. This time-independent equation, because of the stock price moving only between support and resistance, not surprisingly, is a Schrodinger equation for a particle in a box.

Considering the support and resistance levels as walls of a box, the price of the stock matches exactly the move of a particle between the walls. As particle can penetrate out of the box, the stock price could break out of the region bounded by support and resistance too. The breakouts very often occur in the stock market.

To deduce the probability of a stock price penetration out of support or resistance is the same as to calculate the probability of a particle to tunnel out of the box, also called transmission coefficient.

The time-independent equation is used exactly for the purpose of derive the transmission coefficient for the stock price. The probability of price penetration out of the potential wall is deduced applying the same scheme as particle in a box case. It should be noticed here that transmission coefficient is depending only on interest rate, volatility and strike price.

Although, the transmission coefficient formula may look complicated, it is very easy to be solved, and traders could find valuable insights about the best moment to involve themselves in an option contract.

A special attention is dedicated to **λ** constant. The values of **λ** dictate the applicability of transmission coefficient only for stocks having the price raging in an interval from 2 USD to 15 USD. It seems that the smaller the distance between resistance and support level is, the clearer the tunneling effect will be.

Knowing **λ** value another important result can be derived; once the price tunnel through the support or resistance level, one can tell what the stock price will be at the end of the price move, in other words, the magnitude of price penetration.

Traders and investors aware of the **λ** values will better appreciate not only the moment of getting involved in an option contract but, also, the appropriate price of the underlying stock at which the option contract should be exercised.

A third important result formulated in this paper is the prediction of a short term abrupt fall in stock volatility prior to price tunneling out of the range bound.

| Interest rate -r | Volatility -σ | Transmission coefficient- T(%) | Distance of penetration |
|---|---|---|---|
| 0.01 | 0.53 | 73 | 4.88 |
| 0.02 | 0.53 | 75 | 2.75 |
| 0.03 | 0.53 | 79 | 1.80 |
| 0.04 | 0.53 | 83 | 1.24 |
| 0.05 | 0.53 | 87 | 0.85 |
| 0.06 | 0.53 | 91 | 0.57 |
| 0.07 | 0.53 | 95 | 0.35 |

Table 1. Probability of tunneling and distance of penetration for different values of interest rate and constant volatility

| Interest rate -r | Volatility -σ | Transmission coefficient- T(%) | Distance of penetration |
|---|---|---|---|
| 0.05 | 0.43 | 91 | 0.53 |
| 0.05 | 0.53 | 87 | 0.85 |
| 0.05 | 0.63 | 85 | 1.15 |
| 0.05 | 0.73 | 84 | 1.42 |
| 0.05 | 0.83 | 83.95 | 1.67 |
| 0.05 | 0.93 | 83.69 | 1.91 |
| 0.05 | 0.97 | 83.64 | 2.00 |

Table 2. Probability of tunneling and distance of penetration for different values of volatility and constant interest rate

| | date | r | σ | Price at resistance | Price at support | K | d | T | Fall in volatility |
|---|---|---|---|---|---|---|---|---|---|
| LNKD - LINKEDIN CORP | 07-08.02.2013 | 0.03 | 0.47 | 127.2 | 123.3 | 3.9 | 0.058114 | 0.998675 | 0.63 to 0.39 |
| GOOG - GOOGLE INC A | 22-23.01.2013 | 0.03 | 0.15 | 704.7 | 702.6 | 2.1 | 0.136068 | 0.95 | 0.40 to 0.15 |
| HUM - HUMANA INC | 28.03-02.04.2013 | 0.03 | 0.31 | 70.08 | 66.95 | 3.13 | 0.08455 | 0.9948 | 0.43 to 0.25 |
| NFLX - NETFLIX INC | 23-24.01.2013 | 0.03 | 0.55 | 101.17 | 97.81 | 3.36 | 0.921744 | 0.933 | 0.95 to 0.55 |

Table 3. Transmission coefficient and fall in volatility prior price tunneling for some stocks.

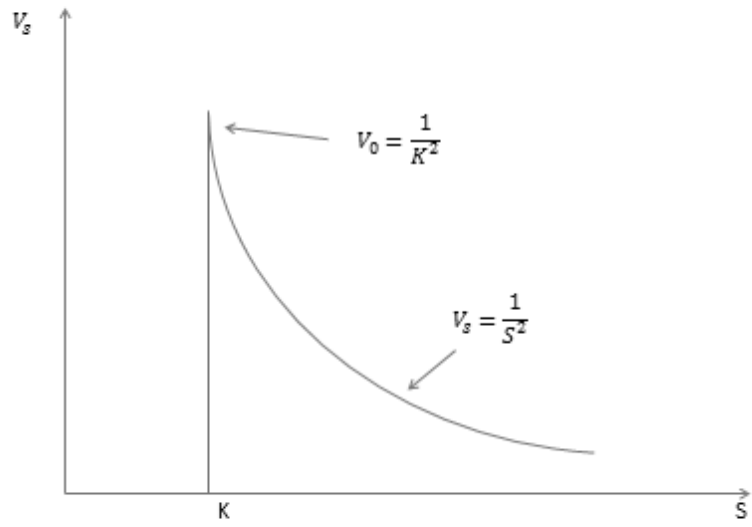

Figure 1. Shape of the stock price potential.

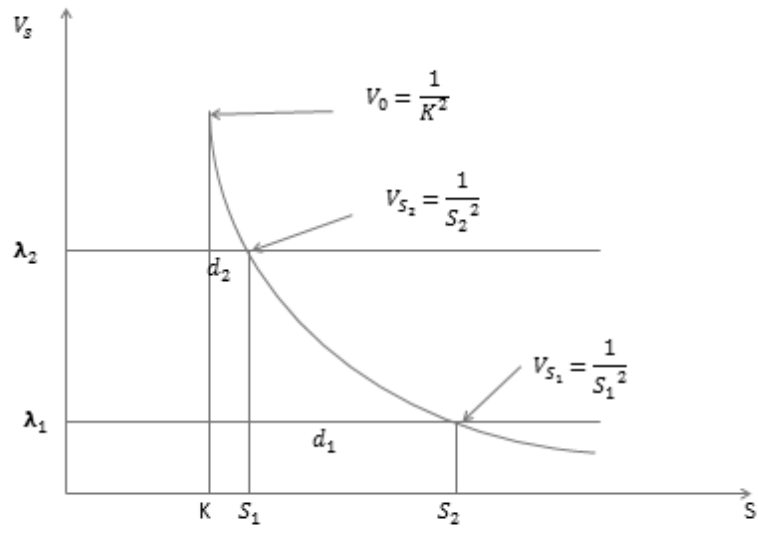

Figure 2. Different levels for **λ** constant.

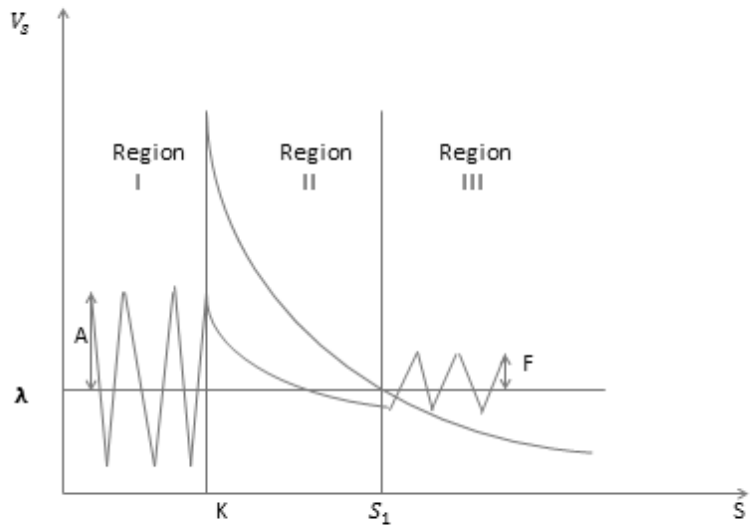

Figure 3. Behavior of $\psi_{(S)}$ function outside and inside the potential wall.

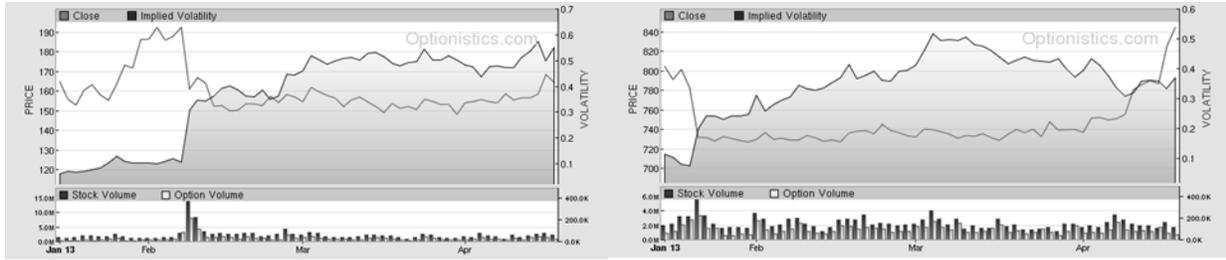

Figure 4. The fall in the stock volatility preceding the tunneling price effect for LNKD(left) and GOOG(right) stocks.